\documentclass[A4,12pt]{article}
\newcommand{\Ra}{\rangle}
\newcommand{\La}{\langle}

\usepackage{amssymb,amsmath,euscript}
\begin{document}
\begin{center}
{\large {\bf  SQUEEZED COHERENT STATES AS THERMAL-LIKE ONES }\\
 \small(ON THE PROBLEM OF INCORPORATING THERMODYNAMICS INTO QUANTUM THEORY)}\\
\vspace{.23cm}
\large {  A.D.Sukhanov$^1$,  O.N. Golubeva$^2$, and V.G.Bar'yakhtar$^3$ }\\
\par\medskip
$^1$\small {Theoretical Physics Laboratory, Joint Institute for
Nuclear Research, Dubna, Russia} \small{ ogol@oldi.ru}\\$^2$
People's Friendship University of Russia, Moscow, Russia.
 \small{ ogol@mail.ru}\\$^3$\small {Institute of Magnetism, National Academy of Sciences of Ukraine, Kiev,
Ukraine}.\small{<victor.baryakhtar@gmail.com>}
\end{center}
\begin{abstract}
In our paper [1], we proposed an original approach to the
incorporation of stochastic thermodynamics into quantum theory. It
is based on the concept of consistent inclusion of the holistic
stochastic environmental influence modeled by arbitrary Bogoliubov
vacua. In our study, we implement a possibility of describing the
states of the system "an object + arbitrary vacuum"\,   based on the
wave function.  The fact that the quantum language is required
becomes especially appreciable in the cases where thermal and
quantum fluctuations occur simultaneously. This allows analyzing
different types of quantum states and estimating the degree of their
applicability to the description of thermal effects. The main result
of our study reduces to the fact that squeezed coherent state (SCS)
cannot be used to consistently explain quantum-thermal effects under
conditions of object equilibrium with the stochastic environment at
any temperatures.
\end{abstract}

\section*{\small 1. Generalization of the concept of thermal equilibrium to quantum phenomena}
As is well known, in the~\emph{standard} thermodynamics,  the
thermal equilibrium state is fixed by the special parameter, namely,
the Kelvin temperature, which does not reach the zero value in
principle. As this takes place, this state satisfies the zeroth law
of thermodynamics, i.e., the assertion that the object temperature
$T$ and the environment temperature $T_0$ are equal (if only on the
average). In this case, it is meant that the model of a classical
thermostat is brought in correspondence with the external
macro-environment. As a result, only thermal fluctuations that are
significant only at high temperatures are considered within the
framework of the~\emph{ standard } thermodynamics. In addition, it
is tacitly taken that the unavoidable quantum stochastic
environmental influence is not taken into account.

At the same time, it is very significant at sufficiently low
temperatures, where it manifests itself through quantum
fluctuations. On the contrary, the concept of thermal equilibrium is
not used in quantum mechanics (i.e., at zero Kelvin temperature),
because the presence of any thermal contact with the environment is
not assumed. Thus, as if the standard thermodynamics and quantum
mechanics are at different poles of the temperature scale.
Nevertheless, it turns out that the concept of thermal equilibrium
can be joined to that of the apparatus of quantum mechanics. The
fact that the thermal equilibrium as a fundamental concept of
equilibrium thermodynamics has the stability property, which is also
inherent in certain quantum states, must be used in this case. It
also must be kept in mind that a peculiar uncertainty relation
between the temperature $T$ and the entropy $S$ [2] in the form
\begin{equation}\label{1-00}
\Delta T\cdot\Delta S \geqslant k_B
\end{equation}
is valid in thermal equilibrium. We note that, at the same time, the
uncertainty relation is traditionally related to the most important
statements of quantum mechanics. In the most general case, it has
the form of the Schr\"odinger uncertainty relation (SUR)
\begin{equation}\label{2}
\Delta p\cdot\Delta q \geqslant\big|\La\psi|\delta {\hat
p}\cdot\delta {\hat q}|\psi\Ra\big|\equiv |\widetilde R_{p\;q}|^2,
\end{equation}
where the notation $\Delta p\equiv\sqrt{\overline{(\Delta p)^2}}$
and $\Delta q\equiv \sqrt{\overline{(\Delta q)^2}}$ is used in the
left-hand side of the inequality, and the variances of the
momentum~$\overline{(\Delta p)^2 }$ and the coordinate~
$\overline{(\Delta q)^2 }$ in the state~$|\psi\Ra$ are calculated in
accordance with the definitions.

The complex quantity in the right-hand side of this expression
\begin{equation}\label{3}
\widetilde{R}_{p\,q}=\langle\Delta p|\Delta
q\rangle=\langle\,|\Delta \widehat{p}\,\Delta
\widehat{q}\,|\,\rangle\end{equation} has a double sense. On the one
hand, it is the amplitude of the transition from the state~$|\Delta
q\rangle $ to the state~ $|\Delta p\rangle $. However, on the other
hand, it can be treated as a mean or the quantum correlator with
respect to the arbitrary state~ $|\; \rangle$  of the operator~
$\mathcal {\widehat{S}_{R}}\equiv\Delta \widehat p\,\Delta
\widehat{q}$, which we called previously [1] the operator of
stochastic influence -- Schr\"odingerian. In this case, the fact
that this quantity is nonzero is an underlying attribute of a
nonclassical theory in which the stochastic environmental influence
on the object plays an important role.

Thus, a certain crossing between the conceptual apparatuses of these
two theories has already been outlined. To extend it, we study the
possibility of bringing the set of quantum states described by wave
functions in correspondence with equilibrium thermal states. To do
this, we proceed from the following considerations.

First of all, we assume that, in the general case, the existence of
inequalities similar to~ (1) in thermodynamics and to~ (2) in
quantum theory is provided by the nonzero value of the correlator
forming their right-hand side. It is also necessary to take into
account that the thermal equilibrium has the stability property. It
can be assumed that the correlation between the corresponding
quantities in thermodynamic inequality~(1) also exists and plays a
certain role in the preservation of the stability of the thermal
equilibrium state. Therefore, it would be natural to begin by
determining a stable \emph{quantum} state that can be assumed to be
adequate to the thermal equilibrium state if the stochastic thermal
influence is taken into account additionally.
Thus, in our opinion, the following conditions must be satisfied in the sought quantum state:\\
a) the correlator of form~ (3) is nonzero;\\ b) the Schr\"odinger
uncertainty relation (SUR) of form~ (2) has the form of the
equality, i.e., becomes saturated, providing the state stability.\\
Of course, condition~b) is reached when the SUR left-hand side
\begin{equation}\label{4}
\Delta p\cdot\Delta q\equiv\mathcal U\mathcal P,
\end{equation}
called the uncertainties product and regarded as a single holistic
quantity takes a value that is as minimum as possible.

To find the general form of the wave function corresponding to the
sought state, we use the Schwartz-von-Neumann theorem [3]. In
accordance with the corollary from this theorem, the equality sign
in SUR~(2) is reached in the state~ $\big|\psi_{\alpha}\Ra$ when the
vectors~ $|\delta p\Ra$ and~$|\delta q\Ra$ are proportional to each
other in the Hilbert space, i.e.,
\begin{equation}\label{5}
\delta{\hat p}\big|\psi_{\alpha}\Ra=(i\gamma\cdot
e^{i\alpha})\delta{\hat q}\big|\psi_{\alpha}\Ra.
\end{equation}
Here, the dimensional parameter~ $\gamma>0$ and the phase
parameter~$\alpha\gtreqqless 0$. To simplify our calculation, we
assume that in this state, the average values of the momentum and
the coordinate are zero so that $\delta {\hat p}= \hat p$ and
$\delta {\hat q}= \hat q$ in (3) and below. From formula~(5), the
equation
\begin{equation}\label{6}
(\hat p-i\gamma e^{i\alpha}\hat q)|\psi_{\alpha}\Ra=0
\end{equation}
can be obtained. It is interesting that it resembles (up to the
normalization factor) the result of the action of the annihilation
operator~$$\hat a= \frac {\hat p - i \zeta\hat
q}{\sqrt{2\hbar\zeta}}$$ on the state~$|\psi_{\alpha}\Ra$.
Expression~(6) in the coordinate representation takes the form of
the differential equation for the unknown
function~$\psi_{\alpha}(q)$:
\begin{equation}\label{7}
\frac\hbar i\frac{d}{dq}\psi_{\alpha}-(i\gamma e^{i\alpha}) q\cdot
\psi_{\alpha}=0.
\end{equation}
Solving it and using the normalization condition, in the general
case, we obtain the complex function
\begin{equation}\label{8}
\psi_{\alpha}(q) =[2\pi (\Delta
q_0)^2\frac{1}{\cos\alpha}]^{-1/4}\exp \left\{-\frac{q^2} {4(\Delta
q_0)^2}e^{i\alpha}\right\}
\end{equation}
as a universal wave function~$\psi_{\alpha}(q)$. Here,
\begin{equation}\label{9}
(\Delta q_0)^2=\frac{\hbar}{2\gamma}.\;\;\;\;\;\;\alpha\ne\frac\pi2
\end{equation}
The physical meaning of the function~ $\psi_{\alpha}(q)$ is
clarified if  the fact that for~$\alpha=0$, Eq. (8) for the
function~$\psi_0(q)$ is equivalent to the equation for the
state~$|0\Ra$ (which is adopted to be called the~ \emph{cold vacuum
}) is taken into account. Accordingly, we assume that for arbitrary
value~$\alpha\ne0$, the state~$\psi_{\alpha}$ describes
an~\emph{arbitrary vacuum }. We note that unlike the approach
proposed above, the standard way of obtaining the state of the
arbitrary vacuum, which we used in~[2], is related to the
application of the Bogoliubov $(u,v)$- transformation. The
coordinate variance~ $\overline{(\Delta q_{\alpha})^2}$ in the
arbitrary-vacuum state~$|\psi_{\alpha}\Ra,$  calculated using wave
function~(8) has the form
\begin{equation}\label{10}
\overline{(\Delta
q_{\alpha})^2}=\int_{-\infty}^{+\infty}\psi^*_\alpha q^2\psi_\alpha
dq=\frac{\hbar}{2\gamma}\cdot\frac{1}{\cos\alpha}
\end{equation}
The momentum variance~$ \overline{(\Delta p_{\alpha})^2}$ and the
"coordinate--momentum"\, correlator~\; $\big|\La\delta
p_{\alpha}|\delta q_{\alpha}\Ra\big|$ are related by the
proportional dependence with the coordinate
variance~$\overline{(\Delta q_{\alpha})^2}$
\begin{equation}\label{11}
\overline{(\Delta p_{\alpha})^2}=\gamma^2(\Delta
q_{\alpha})^2=\frac{\hbar\gamma}{2}\cdot\frac{1}{\cos\alpha};
\end{equation}

\begin{equation}\label{12}
\big|\La\delta p_{\alpha}|\delta q_{\alpha}\Ra\big|=\big|i\gamma
e^{i\alpha}\big|\La\delta q_{\alpha}|\delta
q_{\alpha}\Ra=\gamma(\Delta
q_{\alpha})^2=\frac{\hbar}{2}\cdot\frac{1}{\cos\alpha}.
\end{equation}

Thus, we assume that the function~ $\psi_{\alpha}$ describes
the~\emph{equilibrium} with the arbitrary vacuum. However, the
intuitive assumption of the possibility of describing thermal states
by the functions~ $\psi_{\alpha}$ requires a further substantiation.
To do this, it is necessary (if it is possible) to relate the
obtained expressions to the Kelvin temperature, which has no direct
pre-image in quantum mechanics, and also to establish which values
of the parameter~ $\alpha$ correspond to the thermal states.

\section*{\small 2. Squeezed coherent states of the arbitrary vacuum as thermal-like states}
At this stage, we analyze squeezed coherent state (SCSs) in order to
establish how adequate to the some kind of thermal states they are.
The cold-vacuum states are described by real wave functions and
occur in the case where the parameter~$\alpha$ is zero in
formula~(8). Accordingly, $\cos\alpha=1$ in formulas~ (14) and~(15)
so that the coordinate and momentum variances have the forms
\begin{equation}\label{13}
\overline{(\Delta p_0)^2}= \frac{\hbar\gamma}{2} ;
\end{equation}

\begin{equation}\label{14}
\overline{(\Delta q_0)^2} =\frac{\hbar}{2\gamma}
\end{equation}
In what follows, we establish that such wave functions describing
the cold-vacuum states cannot be interpreted as truly thermal ones.
Expressions~ (13) and~(14) allows obtaining the average values of
the kinetic~$<K>$ and potential~$<\Pi>$ energies of the system under
consideration in the case where $\overline{(p_0)^2}=0$
and~$\overline{(q_0)^2}=0$:
\begin{equation}\label{15}
<K>=\frac12\overline{(\Delta p_0)^2}=\frac{\hbar\gamma}{4},
\end{equation}

\begin{equation}\label{16}
<\Pi>=\frac12\overline{(\Delta q_0)^2}=\frac{\hbar\gamma}{4}.
\end{equation}
Thus, the total energy of the quantum oscillator
\begin{equation}\label{17}
U=<K>+<\Pi>=\frac{\hbar\gamma}{2},
\end{equation}
which corresponds to the zero-oscillation
energy~$\dfrac{\hbar\omega}{2}$ if~$\gamma$ is endowed with the
meaning of the frequency: $\gamma\rightarrow\omega.$ In other words,
the function~$\psi_0$ corresponds to the state at the zero Kelvin
temperature. Of course, it occurs solely under conditions under
which there is no uncontrolled thermal influence, so that it can be
considered as thermal exclusively in the Pickwickian sense.

The problem whether the state~$|\psi_0\Ra$ can nevertheless be
interpreted as an equilibrium one deserves additional discussions.
It is easy to see that on the one hand, it provides the SUR
saturation. But, on the other hand, relation~(2) in this case
transforms into the saturated Heisenberg UR
\begin{equation}\label{18}
\Delta p_0\cdot\Delta q_0=\big|\La\psi_0|\frac 12[\hat p,\hat
q]|\psi_0\Ra\big|=\frac\hbar2.
\end{equation}
Here, $\dfrac\hbar2\equiv\mathbb J_0$ is the measure of the purely
quantum environmental influence. Thus, at~ $ \alpha=0$, each of the
quantities ($\mathcal U\mathcal P$ and the correlator~ $\big|\La
\psi_0|\hat p\cdot\hat|\psi_0\Ra\big|$) have values as minimum as
possible independently. In other words, the state~$|\psi_0\Ra$ can
indeed be regarded as a stable state that is similar to the~
\emph{equilibrium}; but the cold vacuum plays the role of as
"thermostat"\, in this case, because this state corresponds to the
minimum vacuum energy~ $\mathbb U_0=\dfrac{\hbar\omega}{2}$.

Thus, it is improper to bring squeezed coherent states in
correspondence with any equilibrium \emph{thermal} states. Only the
equilibrium with the cold vacuum, which is typical of maximally
isolated systems (i.e., systems experiencing a solely unavoidable
quantum stochastic influence), can be discussed notably arbitrarily.
Probably, it is no mere chance that considering squeezed states,
Umedzawa [4] called them not truly thermal ones but thermal-like
states. Further search for the possibility of bringing the states~
$|\psi_{\alpha}\Ra$ (with the arbitrary value~$\alpha\ne0$) in
correspondence with the thermal states requires studying correlated
coherent states.

\small References

[1] A.D. Sukhanov and O.N. Golubjeva. Arbitrary vacuum as a model of
stochastic  influence of environment: on the problem of
incorporating   thermodynamics into quantum theory. Physics of
Particles and Nuclei Letters, 9(3):303.
 Pleiades Publishing, Ltd.(2012).\\

[2] A.D. Sukhanov and O.N. Golubjeva. Toward a quantum
generalization of equilibrium statistical thermodynamics:
$(\hbar,k)$- dynamics. Theoretical and Mathematical Physics, 160(2):
1177. Springerlink(2009)\\

[3.] J.von Neumann. Mathematical Foundations of Quantum Mechanics. Princeton University Press,1996\\

[4]  H.Umezawa. Advanced Field Theory. Micro-, macro-, and Thermal
Physics. AIP, N.-Y. 1993.

\end{document}